# AstroSeis – A 3-D Boundary element modeling code for seismic wavefields in irregular asteroids and bodies

Yuan Tian[1] and Yingcai Zheng[1]

[1] Department of Earth and Atmospheric Sciences, University of Houston

Correspondence: ytian10@uh.edu

## 1   Abstract

We developed a 3-D elastic Boundary Element Method (BEM) computer code, called AstroSeis, to model seismic wavefields in a body with an arbitrary shape, such as an asteroid. Besides the AstroSeis can handle arbitrary surface topography, it can deal with a liquid core in an asteroid model. Both the solid and liquid domains are homogenous in our current code. For seismic sources, we can use single forces or moment tensors. The AstroSeis is implemented in the frequency domain and the frequency-dependent Q can be readily incorporated. The code is in MATLAB and it is straightforward to set up the model to run our code. The frequency-domain calculation is advantageous to study the long-term elastic response of a celestial body due to a cyclic force such as the tidal force with no numerical dispersion issue suffered by many other methods requiring volume meshing. Our AstroSeis has been benchmarked with other methods such as normal modes summation and the direct solution method (DSM). This open-source AstroSeis will be a useful tool to study the interior and surface processes of asteroids.

## 2 Introduction

Asteroids and meteorites provide clues to understand the formation of planetesimals in the solar system. However, the internal elastic structure of asteroids is poorly constrained (Johansen *et al.*, 2015; Walsh, 2018). Seismology is very useful in imaging the interior of a solid body(Asphaug, 2020). Murdoch *et al.* (2017) showed passive seismic on the asteroid, Didymoon, could be used to distinguish different proposed internal models. The normal mode analysis of rubble-pile asteroids could reveal information about their internal structure (Chujo *et al.*, 2019). On the other hand, many astronomical/planetary processes are related to seismic waves or shaking. For example, seismic shaking can reshape an asteroid surface because they can exert large stresses exceeding the low gravity (Asphaug *et al.*, 1996). The excitation of the asteroid normal modes could change the topography of the asteroid (Quillen *et al.*, 2019). Recently, Tian and Zheng (2019) proposed a tidal-seismic resonance effect where seismic waves in a planet excited by an orbiting moon's tidal force can in turn influence the moon's orbit. Seismology can be used to analyze the asteroid Bennu's OSIRIS-REx images (DellaGiustina et al., 2019) and lidar for clues in features like run-outs, slopes, particle sorting to infer internal compositions and physical properties. The potential seismology data on the asteroids can be obtained by orbital laser vibrometer seismology (Sava and Asphaug, 2019; Courville and Sava, 2020). We can also use seismology to examine the resurfacing event caused by the impact on the Ryugu asteroid (Arakawa et al., 2020). Clearly, the ability to model seismic wavefield in irregular bodies is useful in understanding both the interior and surface processes of celestial bodies.

To model seismic waves in an asteroid, we need to address two issues. First, we must be able to consider surface topography because asteroids are irregular in geometry and topography can influence seismic waves significantly. Secondly, some seismic sources are periodic and long in duration such as the tidal force and because such the ability to model in the frequency domain and to incorporate a frequency-dependent Q is also desired. For these reasons, we developed a 3D frequency-domain elastic boundary element method (BEM). In addition, our BEM can also include a liquid core for a geologically differentiated body.

There are many widely used numerical methods to model seismic wavefields in Earth or other celestial bodies. For a 1-D spherical model, we can use the normal mode method (Ben-Menahem and Singh, 1981; Dahlen and Tromp, 1998; Aki and Richards, 2002). This method is the exact solution to model seismic wavefield. The package, MINEOS (Masters *et al.*, 2011), is a publicly available software for normal mode synthetic seismograms. However, its ability to model waves higher than 166mHz is limited for the Earth. The direct solution method (DSM) (Cummins *et al.*, 1994a; Cummins *et al.*, 1994b; Geller and Takeuchi, 1995; Takeuchi *et al.*, 1996; Kawai *et al.*, 2006) is also a 1-D model besed code which can compute high frequencies such as > 1 Hz. In principle, DSM can also handle irregular topographies (Geller and Ohminato, 1994). However, the code released by Takeuchi (Kawai *et al.*, 2006) can only handle 1-D spherical models. Other numerical methods to model seismic wavefields in a more complex model includes the finite difference method (FD) (Boore, 1972; Fang *et al.*, 2014; Zhan *et al.*, 2014). FD is known to have difficulties in modeling wave scattering by irregular topographies. Recent progress by Zhang *et al.* (2012) added the topography modeling capabilities at local scales however its application in global scales remains to be demonstrated. The spectral element method (SEM) (Komatitsch and Tromp, 1999) is a powerful numerical method in modeling waves in 3-D Earth. However, its domain meshing will need specialized software and training. Both FD and SEM are implemented in the space-time domain by discretizing 3-D space into small grids and time into small marching steps. A frequency-dependent Q is not straightforward to be incorporated in FD and SEM. The grid dispersion can be another issue if the modeled seismic field is long in time duration (e.g., due to cyclic tidal forces). To avoid the grid dispersion in FD and SEM, usually very fine grids and accordingly a small time-step should be used, and it leads to expensive computing. On the other hand, if the source location is changed, we need to compute the wavefield again for FD and FEM. However, it is not true for BEM as it can simultaneously handle multiple sources without much added computation. Note that there is no 'best' modeling method universally. Each modeling method, under certain circumstances, can be more or less advantageous than others, depending on the objectives.

Our BEM is based on the boundary integral equation (e.g., Sánchez-Sesma and Campillo, 1991; Ge *et al.*, 2005; Ge and Chen, 2008; Zheng *et al.*, 2016). BEM only discretizes the model on the boundaries and interfaces, which represents a dimension reduction by one. Therefore,

BEM can be less computationally intensive than the other 3-D numerical seismic modeling method (Stamos and Beskos, 1996; Chaillat *et al.*, 2009). Here, we present our BEM method and the associated code to model seismic wavefield in asteroids and small bodies in space. Our BEM is easy to set up and use and coded in MATLAB.

## 3 Method and boundary integral equations

### 3.1 A solid asteroid with topography

First, we show how to use BEM to model seismic wavefield in a solid asteroid. We assume the asteroid is a homogenous solid body with an irregular boundary. In BEM, we only need to know the wavefield on the boundary and we can then compute the wavefield in the entire model.

The boundary integral equation governs the surface seismic displacement field, **u** for an interior domain $\Omega$ (Figure 1a), reads,

$$\chi(\mathbf{x})u_n(\mathbf{x}) = u_{0_n}(\mathbf{x}) + \iint_S \left[ G_{ni}(\mathbf{x}', \mathbf{x}, \omega) t_i(\mathbf{x}') - u_i(\mathbf{x}') C_{ijkl}(\mathbf{x}') G_{nk,l}(\mathbf{x}', \mathbf{x}, \omega) n_j \right] d\mathbf{x}'^2. \tag{1}$$

In this equation, $S$ is the surface of the elastic body including topography. $\chi(x) = 1$ if $x \in \Omega$ and $\chi(x) = \frac{1}{2}$ if $x \in S$. The surface integral should be understood in the sense of the Cauchy principal value if $x$ is on the boundary. In BEM, $x'$ and $x$ are points on S. $n_j$ is the outward surface normal at $x'$; $C_{ijkl}(x')$ is the elastic tensor at $x'$. Here, we assume the medium is isotropic and there are only two independent Lamé parameters in $C_{ijkl}$. $G_{nk}(x', x, \omega)$ is the Green's function, the displacement wavefield along the k-th direction recorded at $x'$ due to a single-force source at $x$ with the force direction along the n-th direction. $G_{kn,l}(x', x, \omega)$ is the spatial directional derivative of the elastic Green's function with respect to $x'$ along the $l$-th direction in the frequency domain. All subscripts (n, i, j, k, l) in equation (1) take a value of 1, 2, or 3 to indicate the component of the vector/tensor field. Because the surface traction, $t_i(x')$, is zero on the free surface, we can neglect it in equation (1).

In equation (1), $u_{0_n}$ is the incident field. For a single-force source $f$, we can directly use Green's function to calculate the incident field:

$$u_{0_n}(x) = \iiint_\Omega f_i(x')\delta(x' - x_0)G_{ni}(x, x', \omega)dx'^3, \qquad (2)$$

where $\Omega$ is the space enclosed by surface S, $x$ is a point on surface S, $x'$ is a point in $\Omega$. $f_i(x')$ is the single force at $x'$ along the i-th direction. $x_0$ is the location of the source within $\Omega$ (Figure 1a).

If the source is a 3-by-3 moment tensor, $M_{ij}$, the incident field is calculated as follows:

$$u_{0_n}(x) = \iiint_\Omega M_{ij}\delta(x' - x_0)G_{ni,j}(x, x', \omega)dx'^3. \qquad (3)$$

We can solve equation (1) for $u(x)$ on the boundary, $x \in S$. We partition the surface into small triangles. Each triangle is called a boundary element. The $I$-th element is called $\Sigma_I$ (Figure 1a). We assume the seismic wavefield $u(x)$ on each surface element is constant. We can then discretize equation (1) and get a system of linear equations:

$$\left(\tfrac{1}{2}\mathbb{I} + T\right)[u] = [u_0], \qquad (4)$$

$[u]$ is a column vector containing the 3-component surface displacements (i.e., the total field including the incident field and scattered fields) on all the elements. $\mathbb{I}$ is an identity matrix. $[u_0]$ is a column vector containing the incident field on all surface elements excited by a single-force or a moment tensor source calculated using equation (2) or (3).

We define a matrix representing the pair-wise field interaction between elements:

$$T(I, I') = \iint_{\Sigma_{I'}} (\mathbf{x}') C_{ijkl}(\mathbf{x}') G_{nk,l}(\mathbf{x}', \mathbf{x_I}, \omega) n_j d\mathbf{x}'^2. \qquad (5)$$

where $I$ and $I'$ are boundary element indices on $S$. They are also representing row index and column index of $T$ matrix.

In the BEM method, we first get the surface displacement $[u]$ on each element by solving the linear algebraic equation (4). The wavefield at any interior point can be calculated using equation (1) using $\chi(x) = 1$ for any interior point $x$ in $\Omega$.

It is worth noting that the matrix $\left(\frac{1}{2}\mathbb{I} + T\right)$ in equation (4) will be nearly singular at its eigen-frequencies. By incorporating a small constant imaginary part for all the angular frequencies (Bouchon et al., 1989), this issue can be mitigated. Alternatively, we can use the hyper-singular BEM (Zheng et al., 2016) by explicitly considering the traction BEM.

### 3.2  Solid body with a liquid core

In this section, we show the BEM modeling for a solid body with a liquid core (Figure 1b). Because of the liquid core in the solid body, we now have two boundaries, we call the outer free surface as $S_1$, and the interface between liquid and solid as $S_2$. Because we divided the surfaces into discrete triangles or boundary elements, we define the I-th and I'-th boundary element on $S_1$ as $\Sigma_I^{(1)}$ and $\Sigma_{I'}^{(1)}$, respectively. We also define the J-th and J'-th boundary element on $S_2$ as $\Sigma_J^{(2)}$ and $\Sigma_{J'}^{(2)}$ (Figure 1b), respectively.

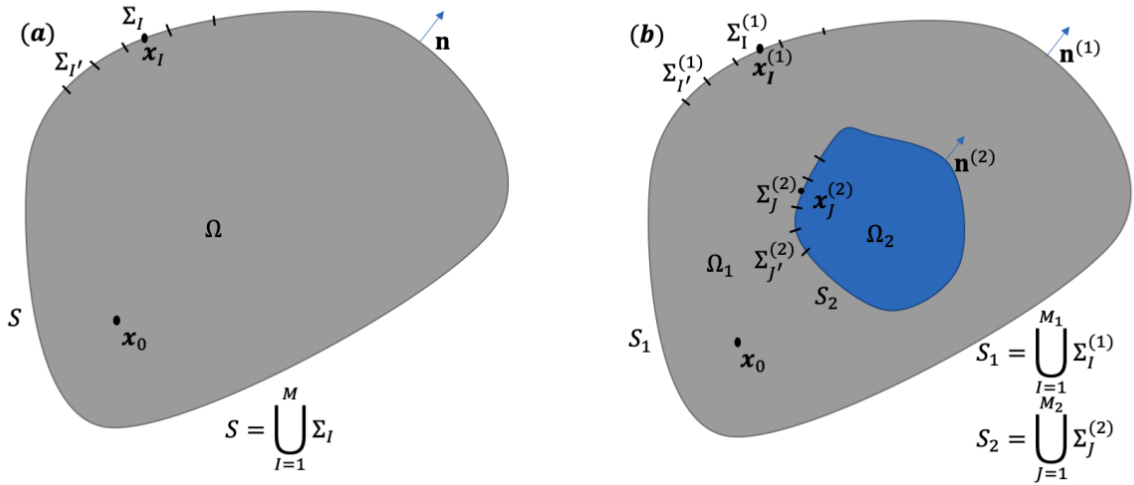

*Figure 1 **Geometry used in BEM. (a) A solid asteroid.** The domain in gray is the solid medium $\Omega$. S is the surface of this domain. We partition the surface into M small triangles. $\Sigma_I$ and $\Sigma_{I'}$ are boundary elements on S with indices $I$ and $I'$, respectively. $x_I$ is the collocation point on the element $\Sigma_I$ which is defined as the center of the inscribed circle on $\Sigma_I$. $x_0$ is the location of the source. $n$ is the surface normal on S. **(b) A solid body with a liquid core schematic.** $\Omega_1$ is representing the solid medium which is the space in gray. $\Omega_2$ is representing the liquid medium which is the space in blue. J and J' are element indices on $S_2$.*

For the seismic wavefield in a liquid medium, the Green's function is simply $G_P = e^{ik_\alpha r}/(4\pi r)$, where $k_\alpha = \omega/v_p$ is the wavenumber in the liquid, $\omega$ the angular frequency, and $r$ the source-receiver distance. We have the boundary integral equation in the liquid medium:

$$\chi(\mathbf{x})P(\mathbf{x}) = P_0(\mathbf{x}) - \iint_{S_2} \left[\frac{\partial G_P(\mathbf{x}',\mathbf{x})}{\partial \mathbf{n}^{(2)}(\mathbf{x}')}P(\mathbf{x}') - G_P(\mathbf{x}',\mathbf{x})\frac{\partial P(\mathbf{x}')}{\partial \mathbf{n}^{(2)}(\mathbf{x}')}\right]d\mathbf{x}'^2, \quad (6)$$

where $\chi(x) = 1$ if $x \in \Omega_2$ and $\chi(x) = \frac{1}{2}$ if $x \in S_2$. $P$ is the total pressure field and $P_0$ is the incident pressure field on the fluid boundary, $S_2$. $x'$ is a point on $S_2$. $x$ can be either on $S_2$ or inside the domain $\Omega_2$. $\mathbf{n}^{(2)}(x')$ is the outward surface normal at $x'$. We can discretize equation (6) using the following matrices (see Figure 1b for the meaning of the symbols):

$$A(J,J') = \iint_{\Sigma_{J'}^{(2)}} \frac{\partial G_P(\mathbf{x}',\mathbf{x}_J)}{\partial \mathbf{n}^{(2)}(\mathbf{x}')}d\mathbf{x}'^2, \mathbf{x}' \in \Sigma_{J'}^{(2)},$$

$$B(J,J') = \iint_{\Sigma_{J'}^{(2)}} G_P(\mathbf{x}',\mathbf{x}_J)d\mathbf{x}'^2, \mathbf{x}' \in \Sigma_{J'}^{(2)}. \quad (7)$$

In the solid domain, we can write the BIE as

$$\chi(x)u_n(x) = u_{0_n}(x) + \iint_{S_1 \cup S_2}\left[G_{ni}(x',x,\omega)T_i(x') - u_i(x')C_{ijkl}(x')G_{nk,l}(x',x,\omega)n_j(x')\right]dx'^2, \quad (8)$$

where $\chi(x) = 1$ for an interior point $x \in \Omega_1$ and $\chi(x) = \frac{1}{2}$ if $x \in S_1 \cup S_2$. In the boundary element method, both $x$ and $x'$ are on $S_1 \cup S_2$. We can discretize the equation (7) using matrices defined below (see Figure 1b for symbols):

$$T^{(11)}(I,I') = \iint_{\Sigma_{I'}^{(1)}} C_{ijkl}(\mathbf{x}')G_{kn,l}(\mathbf{x}',\mathbf{x}_I,\omega)n_j^{(1)}d\mathbf{x}'^2, \mathbf{x}' \in \Sigma_{I'}^{(1)};$$

$$T^{(12)}(I,J') = \iint_{\Sigma_{J'}^{(2)}} C_{ijkl}(\mathbf{x}')G_{kn,l}(\mathbf{x}',\mathbf{x}_I,\omega)(-n_j^{(2)})d\mathbf{x}'^2, \mathbf{x}' \in \Sigma_{J'}^{(2)},$$

$$G^{(12)}(I,J') = \iint_{\Sigma_{J'}^{(2)}} G_{in}(\mathbf{x}',\mathbf{x}_I,\omega)d\mathbf{x}'^2, \mathbf{x}' \in \Sigma_{J'}^{(2)},$$

(9)

$$T^{(21)}(J,I') = \iint_{\Sigma_{I'}^{(1)}} C_{ijkl}(\mathbf{x}')G_{kn,l}(\mathbf{x}',\mathbf{x}_J,\omega)n_j^{(1)}d\mathbf{x}'^2, \mathbf{x}' \in \Sigma_{I'}^{(1)},$$

$$T^{(22)}(J,J') = \iint_{\Sigma_{J'}^{(2)}} C_{ijkl}(\mathbf{x}')G_{kn,l}(\mathbf{x}',\mathbf{x}_J,\omega)(-n_j^{(2)})d\mathbf{x}'^2, \mathbf{x}' \in \Sigma_{J'}^{(2)},$$

$$G^{(22)}(J,J') = \iint_{\Sigma_{J'}^{(2)}} G_{in}(\mathbf{x}',\mathbf{x}_J,\omega)d\mathbf{x}'^2, \mathbf{x}' \in \Sigma_{J'}^{(2)}.$$

We can get the final form of discretized boundary integral equations system for solid-liquid core model by combining BIE in both liquid and solid medium:

$$\frac{1}{2}[u^{(1)}] = [u_0^{(1)}] - T^{(11)}[u^{(1)}] - T^{(12)}[u^{(2)}] + G^{(12)}[t^{(2)}],$$

$$\frac{1}{2}[u^{(2)}] = [u_0^{(2)}] - T^{(21)}[u^{(1)}] - T^{(22)}[u^{(2)}] + G^{(22)}[u^{(2)}],$$

(10)

$$\frac{1}{2}[P] = [P_0] - A[P] + B[q],$$

where $[u^{(1)}]$ is a vector containing the 3-component surface displacements for the elements on $S_1$. Similarly, $[u^{(2)}]$ is a vector containing the 3-component surface displacements for the elements on $S_2$, $[t^{(2)}]$ is the column traction vector on $S_2$. $[u_0^{(1)}]$ and $[u_0^{(2)}]$ are vectors containing the incident field for the elements on $S_1$ and $S_2$, respectively. $[P]$ is a vector containing the pressure field for the elements on $S_1$. $[P_0]$ is a vector containing incident pressure field for the elements on $S_2$. By including boundary condition on the solid-liquid boundary:

$$\frac{\partial P(\mathbf{x})}{\partial \mathbf{n}^{(2)}(\mathbf{x})} = \rho\omega^2 \mathbf{u}^{(2)}(\mathbf{x}) \cdot \mathbf{n}^{(2)}(\mathbf{x}), \mathbf{x} \in S_2,$$

$$\mathbf{t}^{(2)}(\mathbf{x}) = P(\mathbf{x})\mathbf{n}^{(2)}(\mathbf{x}), \mathbf{x} \in S_2,$$

(11)

where $\mathbf{n}^{(2)}(x)$ is the surface normal at $x$ on $S_2$, $\mathbf{t}^{(2)}$ is the surface traction in the solid region on $S_2$. We can now solve equation (10) for $\mathbf{u}^{(1)}, \mathbf{u}^{(2)}$, and $P(x)$ which are the field values on the boundaries. Finally, we can use equations (6) and (8) to calculate the displacement wavefield at any interior point in $\Omega_1$ or $\Omega_2$.

For BEM modeling mesh building, we first partition the surface by creating an approximately uniform triangular tessellation on a unit sphere by minimizing generalized electrostatic potential energy of a system of charged particles using the code by Semechko (2015). We can magnify or shrink the unit sphere mesh to any size we need. We can also directly add the height of the topography to the vertex of the mesh to achieve topography on our model. To implement surface integration on a boundary element (e.g., equations (7) and (9)), we use quadrature integration using a MATLAB program from Xiao and Gimbutas (2010).

## 4 Benchmarking

### 4.1 Benchmark example 1 - Homogenous solid sphere with a single force source

First, we benchmarked the seismic wavefield due to a single force source in a 3D homogenous elastic and spherical solid. We set the compressional wave velocity as $v_p = 6 km/s$ and the shear wave velocity $v_s = 3$ km/s. We set the density as $3000\ kg/m^3$. The spherical body is 20km in radius without topography. The single force source is given as $f = [1,1,1]N$. The location of the source is at depth of 10km, 180° in longitude and 0° in latitude (Figure 2a). We also included attenuation by applying Q values for the P and S wavenumbers,

$$k_\alpha = \frac{\omega}{v_p}\left(1 + \frac{i}{2Q_p}\right),$$
$$k_\beta = \frac{\omega}{v_s}\left(1 + \frac{i}{2Q_s}\right),$$
(12)

where $\omega$ is the angular frequency. The S-wave attenuation is given as $Q_s = 200$ and the P wave attenuation is given as $Q_p = 2.5Q_s$. To avoid the wrap-around effect (Bouchon *et al.*, 1989), we added an imaginary part to the angular frequency, $\omega \to (\omega + \frac{i}{T_m})$, where $T_m$ is the duration of the recording.

We generate a mesh for BEM for the spherical elastic body with a radius of 20km (Figure 2b).

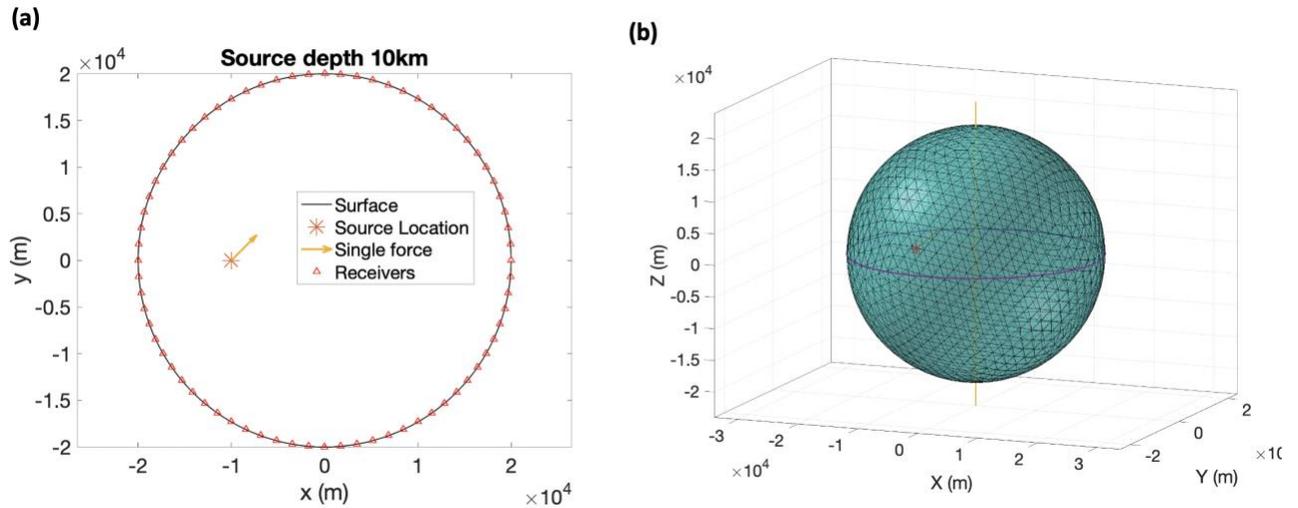

*Figure 2. **A Homogenous model with a single-force source**. **(a)** The source and receivers are on the equatorial plane. The solid body is 20km in radius. We placed the source (indicated by a red star) at depth of 10km, $180°$ in longitude and $0°$ in latitude. The direction of the force is given by the yellow arrow. All the receivers are placed on the equator, spaced at an interval of $5°$. **(b)** 3D view of the boundary mesh. The red star is the source location. The blue circle in the middle is the equator. The yellow line is the axis through the poles.*

Using our AstroSeis code (see Appendix for the usage of our BEM codes), we can model the surface seismic displacement. We compared it with that of the normal mode summation method for the single force source. We found the root-mean-squares (RMS) error of the waveform difference is only 0.88% (Figure 3).

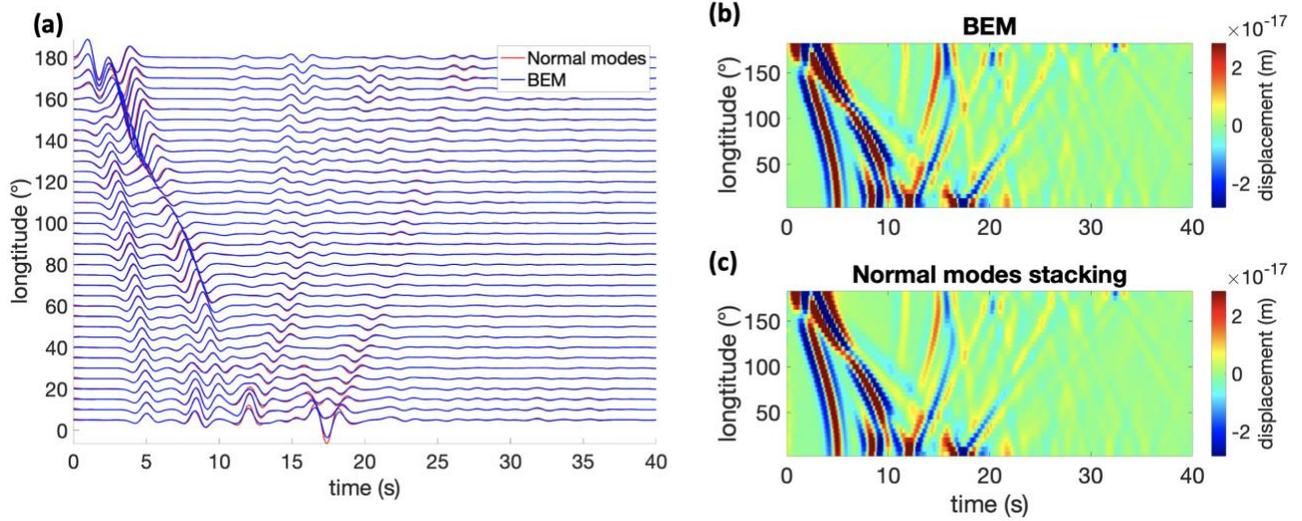

*Figure 3 **Seismic displacement (vertical component) waveform of the model in Figure 2 using two different methods: BEM and normal modes summation, for a single force source**. (a) Wiggle-to-wiggle comparison of seismic waveforms due to a single force source shown in Figure 1. **(b)** A common source gather by BEM; **(c)** Common source gather by the normal mode summation method.*

4.2   Benchmark example 2- Homogenous solid sphere with explosion source

To test whether our method can work with an explosion source, we used the same model as Figure 2. We only changed the source to an explosion source (Figure 4). The moment tensor of the explosion source is given as $M = \begin{bmatrix} 1 & 0 & 0 \\ 0 & 1 & 0 \\ 0 & 0 & 1 \end{bmatrix} N \cdot m$. By comparing the BEM result with the normal mode summation method, the RMS error of the waveform difference is only 0.48% (Figure 5).

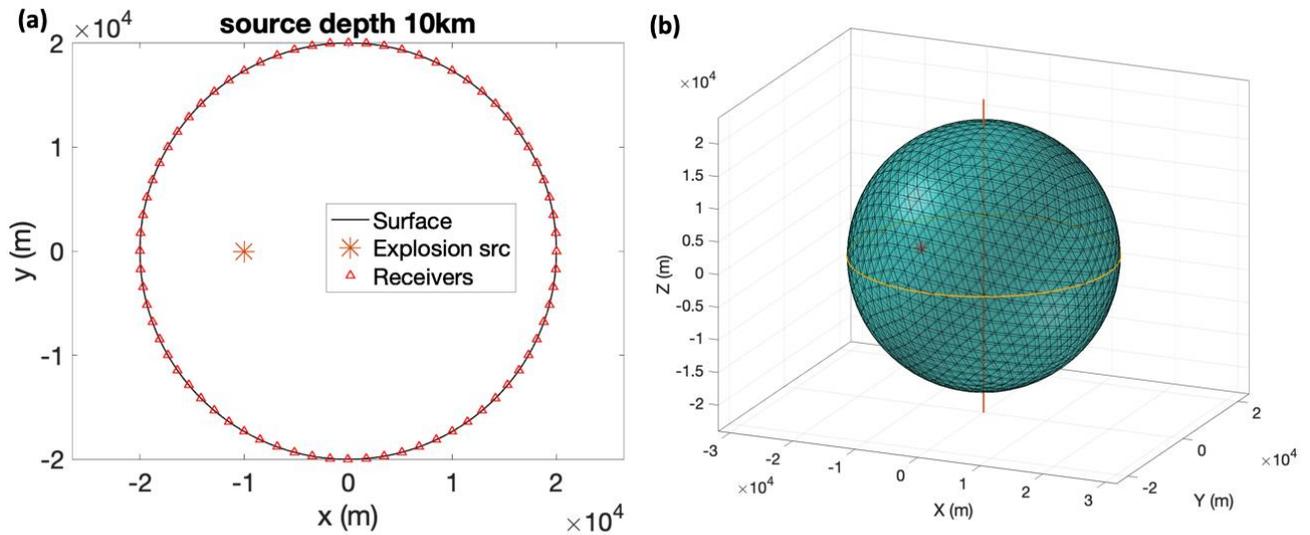

*Figure 4* ***An explosion source in a homogeneous model***. *(**a**) The source and receivers are on the equatorial plane. The solid body is 20km in radius. We placed the source indicated by a red star at depth of 10km, $180°$ in longitude and $0°$ in latitude (on the equator plane). All the receivers are also placed on the equator, spaced at an interval of $5°$. (**b**) 3D view of the boundary mesh. The red star is the location of the source. The orange circle is the equator. The red line is the axis through the poles.*

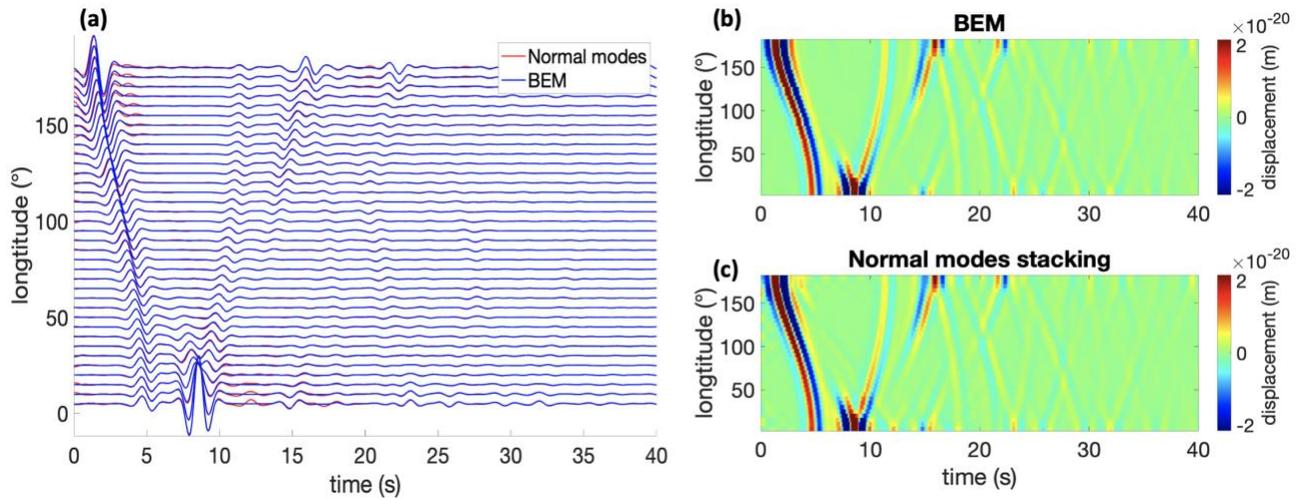

*Figure* 5 ***Seismic displacement fields (vertical component) for the model in Figure 4* computed by *BEM and the normal mode summation for an explosion source*. (a)** W*iggle-to-wiggle comparison of seismic waveforms due to an explosion source shown in Figure 4.* **(b)** *Common source gather computed by BEM;* **(c)** *Common source gather by the normal mode summation method.*

4.3 Benchmark example 3 - Solid sphere with a liquid core

In the third example, we benchmarked our code for a liquid core model (Figure 6). The source is the same explosion source used in example 2. The solid medium is the same as the model in example 1 and 2. The liquid part has a compressional wave velocity of $8km/s$, and density is $4000kg/m^3$. We calculated seismic wavefields using our BEM and DSM. They also showed a good agreement and the RMS error of the waveforms difference between the two methods is 1.45% (Figure 7).

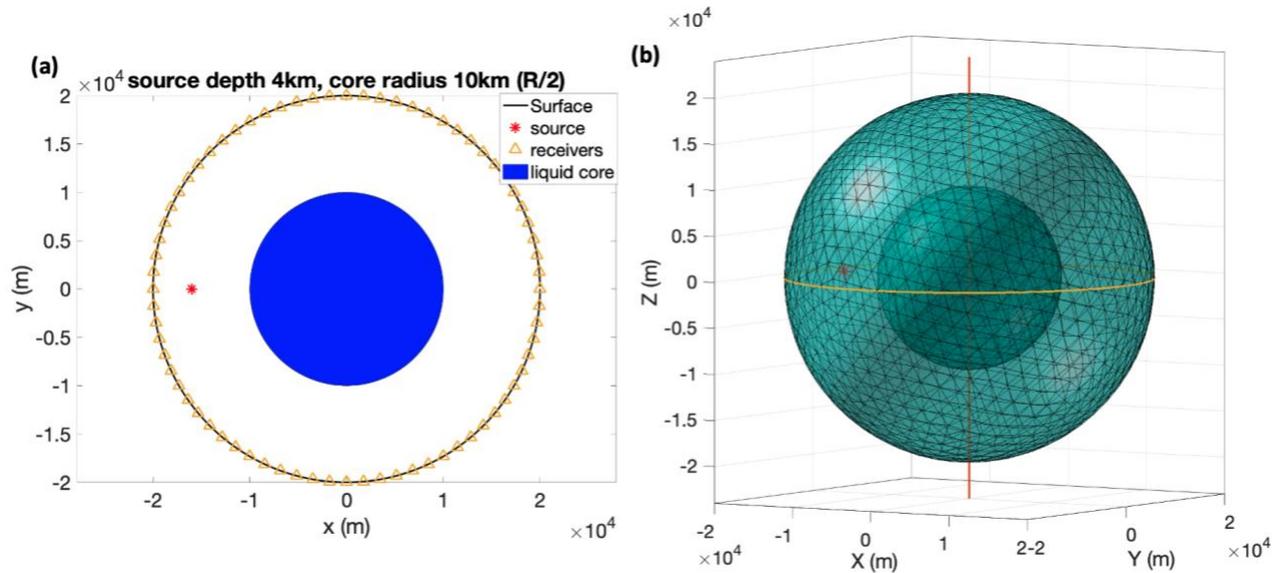

*Figure 6* ***Solid sphere with a liquid core model with an explosion source***. ***(a)*** The *source (red star) and receivers (orange triangles) shown on the equatorial plane. The solid body is 20km in radius. The liquid core is 10km in radius. We placed the source at depth of 4km,* 180° *in longitude and* 0° *in latitude (on the equator plane). All the receivers are on the equator, spaced at an interval of* 5°. ***(b)*** *3D view of the boundary mesh. The red star is the location of the source. The orange circle is the equator. The red line is the axis through the poles.*

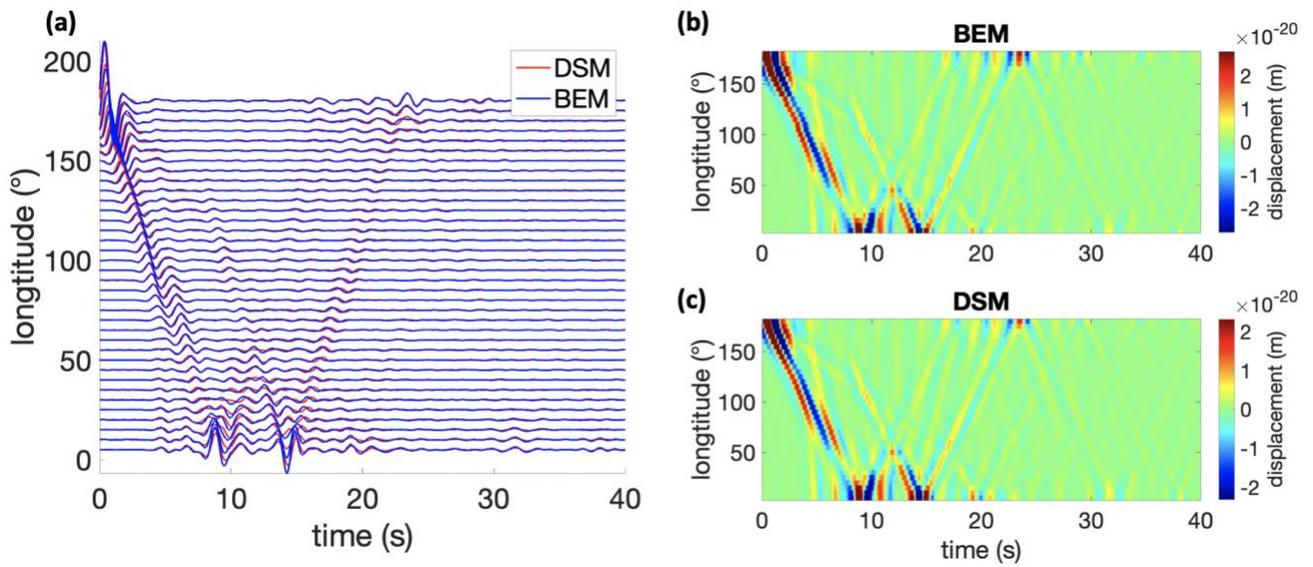

*Figure 7 **Seismic displacement (vertical component) comparison between BEM and DSM for an explosion source**. **(a)** Wiggle-to-wiggle comparison of seismic waveforms (BEM versus DSM). **(b)** Common source gather by BEM; **(c)** Common source gather by DSM (Kawai et al., 2006).*

## 5 Numerical examples for complex models

Here, we show some numerical examples we can compute with our codes. First, we show that our code can model seismic wavefield for a body with a liquid core at an arbitrary location. Second, we can model seismic wavefield in Phobos with its real topography (Willner *et al.*, 2014).

### 5.1 Shifted-core model

We can use our BEM code to model the seismic wavefields in two models: a solid body with a liquid core (the centered-core model), and a solid body with a shifted core. In the shifted-core model, the core is shifted along the y-direction by 2km from the center (Figure 8a). We use AstroSeis to compute seismic displacement (vertical component) wavefields in these two models (Figure 8b). We can clearly see a lack of focusing for the seismic field at the antipode caused by the shift of the core (Figure 8b-c).

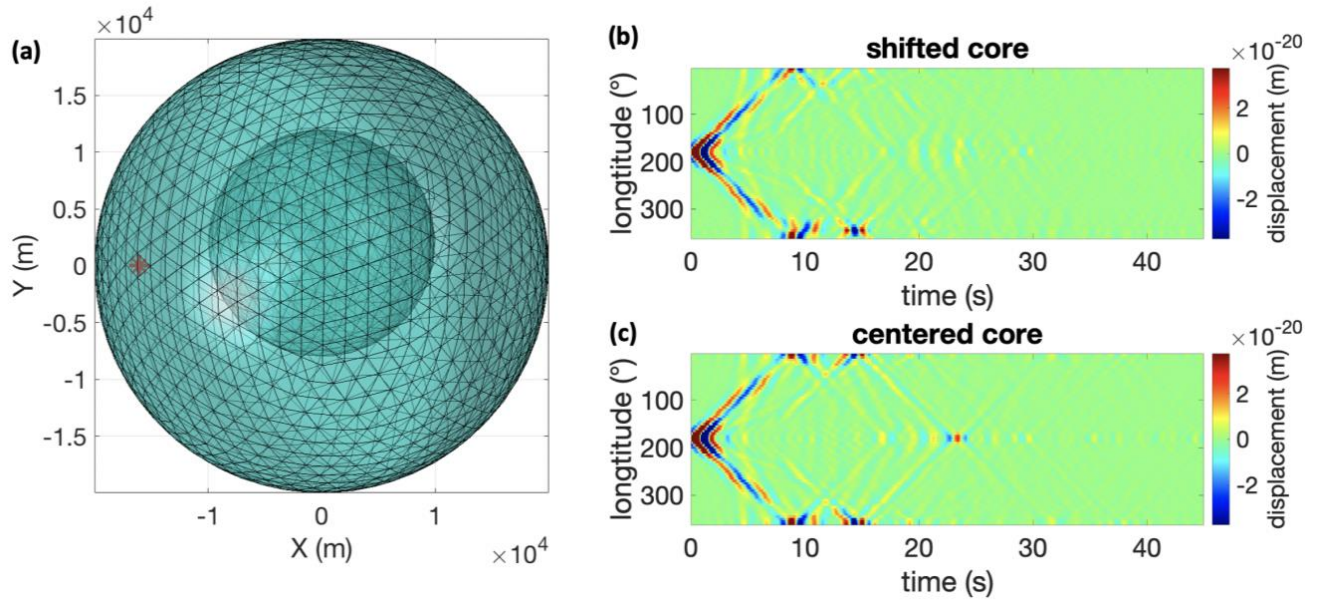

Figure 8 **The seismic wavefield (vertical displacement) in a solid sphere with a shifted liquid core**. (a) The mesh of a solid sphere with a shifted liquid core to calculate seismic displacement wavefield. The core is shifted along the y direction by 2km. The explosion source is given as

$$M = \begin{bmatrix} 1 & 0 & 0 \\ 0 & 1 & 0 \\ 0 & 0 & 1 \end{bmatrix} N \cdot m.$$ The source location is indicted by a red star, at a depth of 4km, 180°

in longitude, and 0° in latitude (on the equator plane). All the receivers are placed on the equator at an interval of 5°. In this solid medium, we set the compressional wave velocity $v_p = 6km/s$ and the shear wave velocity $v_s = 3$ km/s, the density $\rho = 3000\ kg/m^3$; in the liquid medium, we set compressional wave velocity $v_p = 8000m/s$, density $\rho = 4000kg/m^3$. **(b)**

Common source gather for the shifted-core model in (a). **(c)** Common source gather from a solid sphere with a centered liquid core (i.e., no shift).

## 5.2    Seismic modeling for Phobos

In this example, we chose to model seismic fields in Phobos, the closer moon of Mars. Phobos has a very irregular topography (Figure 9a). Next, we compute the seismic displacement wavefield of Phobos with an explosion source at depth of 4km on the Phobos' equator plane (Figure 9). To see how topography modifies the seismic wavefield, we also compute the seismic field for a homogenous spherical model of similar size. We can see the vertical component seismic displacement wavefield has been changed by the topography greatly.

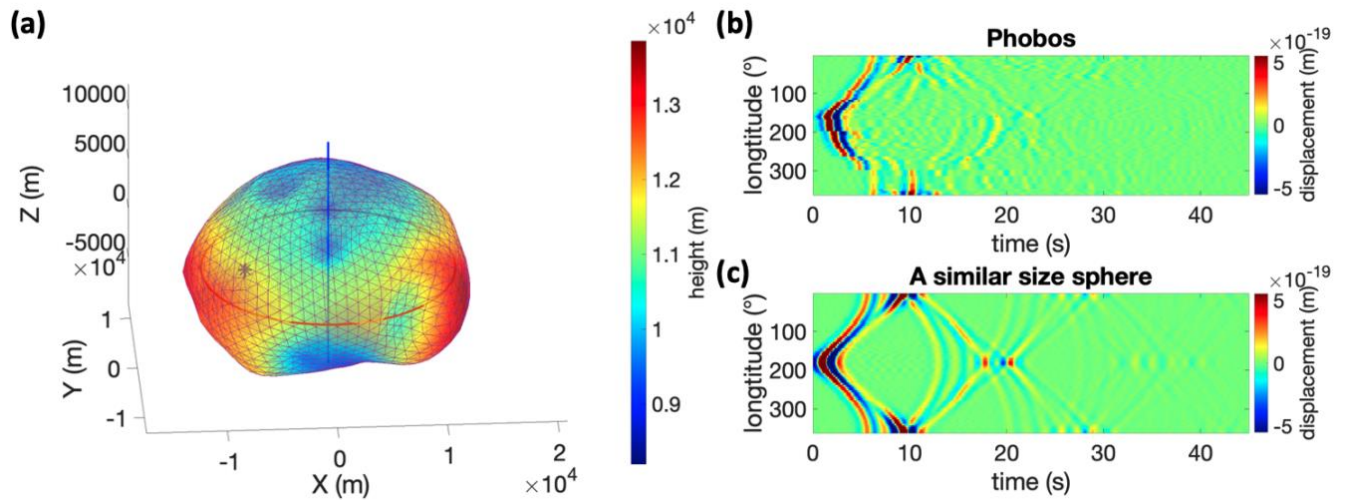

*Figure 9 **Seismic wavefield in Phobos.** (a) The topography model of Phobos with the surface mesh. We define the positive x-axis is $0°$ in longitude, the positive y-axis is $90°$ in longitude. The*

*source location is indicated by a blue star, at depth of 4km, 180° in longitude and 0° in latitude (on the equatorial plane). All the receivers are placed on the approximated equator with an interval of 5°. In this model, we set the compressional wave velocity $v_p = 3km/s$ and the shear wave velocity $v_s = 1\ km/s$, the density $\rho = 1880\ kg/m^3$. And the explosion source is given as*

$$M = \begin{bmatrix} 1 & 0 & 0 \\ 0 & 1 & 0 \\ 0 & 0 & 1 \end{bmatrix} N \cdot m.$$ *(b) Common source gather from the model shown in (a). (c) Common source gather from an elastic sphere with 10.9km in radius.*

## 6   Computational performance

AstroSeis is a frequency domain method and the main computational task in AstroSeis is to assemble the matrices by calculating the surface integration (see equations (5) and (9)) for each frequency. The integration is done using quadrature integration (Xiao and Gimbutas, 2010). To accelerate the performance of the AstroSeis, we parallelize the code using MATLAB "parfor" command to parallelize boundary element integration and the vectorization of the matrix operations. We use an Intel i9-9880H processor with 8 cores on a Macbook to do the computation.

For the Phobos model, we mesh the Phobos surface into 2784 triangles. We take 25 quadrature points in each triangle to calculate the  surface integration. The computational time for one frequency takes about 180 seconds. The total computation depends on the number of the frequencies needed for the modeling. For the Phobos case, the total time we want to model is $T_0 = 50s$ and the maximuml frequency we want to model is $f_{max} = 0.9Hz$, the number of frequency need to be computed is $n_f = f_{max}T_0 = 45$. Therefore, the total computation time for the Phobos modeling is about 8100 seconds using a Macbook. For surface meshing, the number of the triangles depends on the maxium frequency in the modeling. Becausee the S-wave velocity is $v_s = 3 km/s$ in this model, the average size of the triangle should be no larger than the minimal wavelength $\lambda_{min} = v_s/f_{max} = 3333 m$. Therefore, we choose to mesh the Phobos model into 2784 triangles with the maxium triangle size is 3276m. We have tested the modeling result with various triangle sizes and found the results tend to be stable if we choose the grid size smaller than the minimal wavelength $\lambda_{min}$.

For the solid body with liquid core model, we mesh the surface into 3136 triangles and the core interface into 1536 triangles. We used 25 quadrature points for integration. The computational time is about 600 seconds for each frequency. The total calculation time is 27000 seconds on the Macbook mentioned above.

## 7  Conclusions

We have presented the theory and developed a numerical seismic modeling package, AstroSeis, based on our boundary element method. It can handle complex arbitrary surface topography, solid-liquid interfaces, frequency-dependent seismic attenuation, and various source types such as a single force or a moment tensor source. We have verified the validity of our code with the analytical solution (normal mode summation) for a homogenous solid model. We also benchmarked our code against DSM for modeling seismic waves for a liquid core model. We showed the capability of our code in modeling seismic waves in Phobos. We expect our code to be a useful tool in future seismic exploration for asteroids and other planets.

## 8 Data and Resources

The topography data of Phobos comes from Willner *et al.* (2014).The program along with the documentation can be downloaded from: https://github.com/ytian159/AstroSeis.

Takeuchi, N., R. J. Geller, and P. R. Cummins (1996), Highly accurate P–SV complete synthetic seismograms using modified DSM operators, *Geophysical Research Letters*, **23**(10), 1175-1178.

Tian, Y., and Y. Zheng (2019), Rapid falling of an orbiting moon to its parent planet due to tidal-seismic resonance, *Planetary and Space Science*, 104796.

Walsh, K. J. (2018), Rubble Pile Asteroids, *Annual Review of Astronomy and Astrophysics*, **56**(1), 593-624.

Willner, K., X. Shi, and J. Oberst (2014), Phobos' shape and topography models, *Planetary and Space Science*, **102**, 51-59.

Xiao, H., and Z. Gimbutas (2010), A numerical algorithm for the construction of efficient quadrature rules in two and higher dimensions, *Computers & mathematics with applications*, **59**(2), 663-676.

Zhan, X., X. Fang, R. Daneshvar, E. Liu, and C. E. Harris (2014), Full elastic finite-difference modeling and interpretation of karst system in a subsalt carbonate reservoir, *Interpretation*, **2**(1), T49-T56.

Zhang, W., Z. Zhang, and X. Chen (2012), Three-dimensional elastic wave numerical modelling in the presence of surface topography by a collocated-grid finite-difference method on curvilinear grids, *Geophysical Journal International*, **190**(1), 358-378.

Zheng, Y., A. Malallah, M. Fehler, and H. Hu (2016), 2D full-waveform modeling of seismic waves in layered karstic media, *Geophysics*, **81**(2), T25-T34.

# Appendix

## AstroSeis program usage

**Computation**

The program we coded is written in MATLAB, the main script is called: "AstroSeis". All the parameters are in the input file named: "BEM_para". To run the seismic modeling, we can just run "`AstroSeis BEM_para`" in MATLAB (Figure A1).

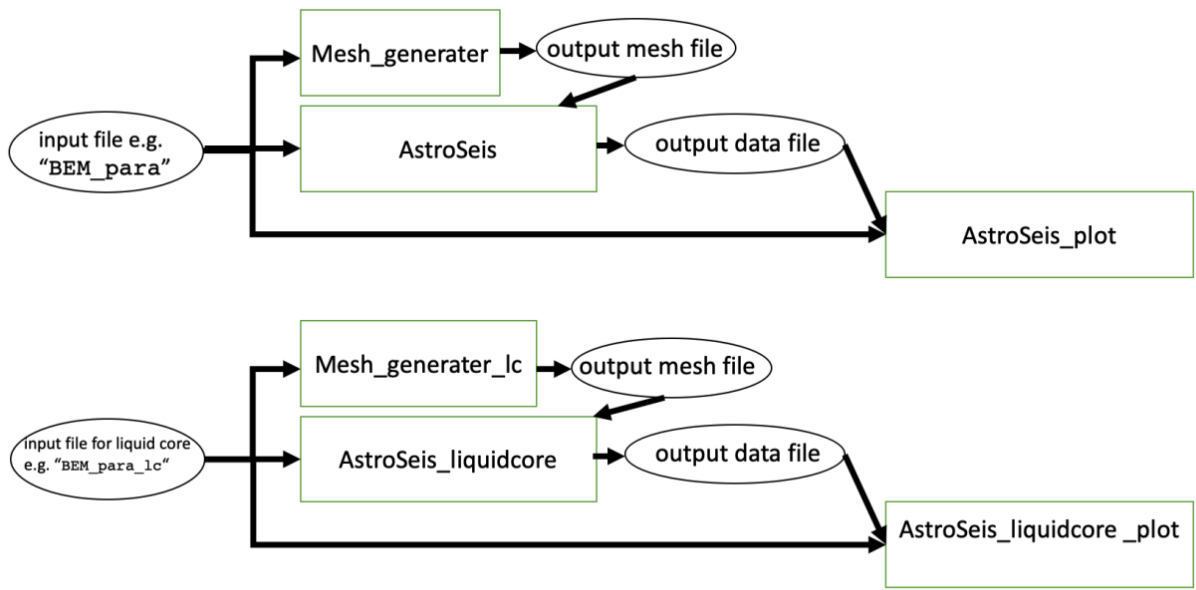

*Figure A1* **Flowchart of AstroSeis.**

As an example (Phobos example), for single force source in homogenous solid model, we can write the input file of this example as the default input file:

The input file, "`BEM_para`", has the following lines

```
Line-1:   phobos_model_20km.mat    # mesh file name
```

```
Line-2:   10993 300  1 # radius of reference sphere, nmesh (number of patch~4*nmesh),
          topography fold
Line-3:   my_mesh.mat # name of the mesh generated from above parameters
Line-4:   demo_out_put.mat  #output wavefield file name
Line-5:   3000 1000 1880 200 # vp, vs, rho:density, Q:attenuation factor
Line-6:   500 0.1 0.5  # nt:number of sampling points , dt:time interval, f0:center
          frequency
Line-7:   moment  0 #type of source could be single or moment, scale of the source
Line-8:   1 1 1 #  fx: x component force, fy: y component force, fz: z component force
Line-9:   1 0 0 1 0 1 # Mxx,Mxy,Mxz,Myy,Myz,Mzz
Line-10:  4e3 0 180 # source depth, latitude, longitude
```

In the input file, all the comment is following after "#".

**Line-1: mesh file name.**

**Line-2: Mesh parameters**: parameters generating a new mesh including the radius for the reference sphere, number of the patches and the fold of topography. The default topography is coming from the Phobos data (Willner *et al.*, 2014). If we set the topography fold as "0", the model will just be a sphere.

**Line-3: New mesh file name**: file name for the newly generating mesh file.

**Line-4: Output wavefield file name.**

**Line-5: Model parameters:** the model parameters including P-wave velocity ($v_p$), S-wave velocity ($v_s$), density ($\rho$) and attenuation factor ($Q_s$).

**Line-6: Time parameters:** parameters like number of sampling points (nt), time interval (dt) and center frequency ($f_0$).

**Line-7: Source parameters**: type of the source and scale of the source. The type of source could be "moment" or "single" for moment tensor source and single force source.

**Line-8: Single force parameters**: three component of single force source if source type is "single". The unit is Newton (N).

**Line-9: Moment tensor parameters**: six component of moment tensor source if source type is "moment". Unit is N-m.

**Line-10**: **Source location parameters**: the location of the source (depth, latitude, longitude)

The default values are for the explosion source. For single force source modeling, we can just change type of source to "single".

For the seismic wavefield in a solid body with liquid core (Benchmark Example-3 in the paper), we can use the MATLAB script "AstroSeis_liquidcore" with the input file "BEM_para_lc":

```
Line-1:   mesh_20km_2_layer.mat   # mesh file name
Line-2:   20000 400  0 # radius of surface, nmesh (number of patch~4*nmesh), topography
          fold
Line-3:   10000 100  0 # radius of solid-liquid boundary, nmesh , topography fold
Line-4:   my_mesh_lc.mat # name of the mesh generated from above parameters
Line-5:   demo_out_put_lc.mat   #output data file name
Line-6:   6000 3000 3000 200 # outer layer: vp, vs, rho:density, Q:attenuation factor
Line-7:   8000    0 4000 200 # core: vp, vs, rho:density, Q:attenuation factor
Line-8:   500 0.1 0.5  # nt:number of sampling points , dt:time interval, f0:center
          frequency
Line-9:   moment  0 #type of source could be single or moment, scale of the source
Line-10: 1 1 1 #  fx: x component force, fy: y component force, fz: z component force
Line-11: 1 0 0 1 0 1 # Mxx,Mxy,Mxz,Myy,Myz,Mzz
Line-12: 4e3 0 180 # source depth, latitude, longitude
```

In this input file, we kept the format as the same, it only has two more lines for the additional layer of the mesh (line 3) and velocity model (line 7). We can run "`AstroSeis_liquidcore BEM_para_lc`" in MATLAB to start modeling (Figure A).

**Visualization**

Visualization is also embedded in this code. The input file can be the same as the computation part. If the computation is finished, we can simply run "`AstroSeis_plot BEM_para`" to do the plotting of the homogenous model (Figure A). For the liquid core model, the input file is similar: "`AstroSeis_liquidcore_plot BEM_para_lc`". We should be able to see the result shortly.